\documentclass[conference,letterpaper]{IEEEtran}

\usepackage[utf8]{inputenc} 
\usepackage[T1]{fontenc}
\usepackage{url}
\usepackage{ifthen}
\usepackage{cite}
\usepackage[cmex10]{amsmath} 

\usepackage{epsfig}
\usepackage{amssymb,amsmath}
\usepackage{mathrsfs}
\usepackage{graphicx}
\usepackage{multirow}
\usepackage{color}
\usepackage{url}

\newcommand{\remove}[1]{}

\newcommand{\X}{\mathcal{X}}
\newcommand{\Y}{\mathcal{Y}}
\newcommand{\bt}{\mathcal{F}_{\mathrm{BIT}}} 
 
\newtheorem{definition}{Definition}
\newtheorem{theorem}{Theorem}
\newtheorem{lemma}{Lemma}

\newtheorem{corollary}{Corollary}

\begin{document}
\title{Non-malleable Coding for Arbitrary Varying Channels\thanks{The author list is alphabetic order}} 




\author{%
  \IEEEauthorblockN{Fuchun Lin}
  \IEEEauthorblockA{Division of Mathematical\\ Sciences, SPMS\\ 
  Nanyang Technological \\ University, Singapore}
    \and
  \IEEEauthorblockN{San Ling}
 \IEEEauthorblockA{Division of Mathematical\\ Sciences, SPMS\\ 
  Nanyang Technological\\ University, Singapore}
    \and
  \IEEEauthorblockN{Reihaneh Safavi-Naini}
  \IEEEauthorblockA{Department of Computer\\  Science\\ University of Calgary\\ Canada}    
    \and
  \IEEEauthorblockN{Huaxiong Wang}
 \IEEEauthorblockA{Division of Mathematical\\ Sciences, SPMS\\ 
  Nanyang Technological\\ University, Singapore}
}


\maketitle


{

\begin{abstract} 
Non-malleable codes protect against an adversary who can  tamper with the coded message by using a tampering function in a specified function family, guaranteeing that the tampering result will only depend on the chosen function and not the coded message.
The codes have been motivated  for  providing protection against tampering with hardware that stores the secret cryptographic keys,  and have
found significant attention in cryptography.
Traditional Shannon model of communication systems  assumes the communication channel is perfectly known to the  transmitter and the receiver.
Arbitrary Varying Channels (AVCs) remove this assumption and have been used to model adversarially controlled channels.
Transmission over these channels has been originally studied with the goal of recovering the sent message, and more recently with the goal of detecting
 tampering with the sent messages. 
In this paper  we introduce non-malleability as the protection goal  of message transmission over these channels, and study binary (discrete memoryless) AVCs  where possible tampering  is modelled by the set of channel states. 
Our main result is that non-malleability for these channels is achievable at a rate asymptotically approaching $1$. 
We also consider the setting of an AVC with a special state $s^{*}$, and the additional requirement that  the message  must be recoverable 
if $s^{*}$ is applied to all the transmitted bits. 
 We  give the outline of a message encoding scheme that in addition to non-malleability, can provide recovery for  all $s^{*}$ channel.

\remove{Non-malleable codes are randomized codes that protect  coded messages against modification by functions in a tampering  function class. These codes are motivated by providing tamper resilience in applications where a cryptographic secret  is stored in a tamperable storage and the protection goal is to ensure that the adversary cannot manipulate the protected message in a message-specific way. In this paper we consider non-malleable codes for protection of robust communication in the Arbitrarily Varying Channel (AVC) model. We consider an AVC application scenario where not only the error correction capacity of the AVC is zero but also that authenticated transmission is impossible. On the other hand, we assume that the adversary might be absent, in which case one fixed state is repeated all through the transmission. We propose a non-malleability coding goal in this scenario that guarantees robust message transmission when the adversary is absent and when the adversary is present, prevents the adversary from tampering the message in a message-specific way. We show that this coding goal can be achieved at a rate as high as the capacity of the channel corresponding to the fixed state when the adversary is absent. We also provide an explicit construction that is sub-optimal in rate.
We discuss our results and directions for future work.
}
\end{abstract}

\section{Introduction}
Reliable message transmission over a noisy channel is a central problem in 
 communication theory.
 Shannon \cite{S48} modelled a message transmission system and, 
  using random coding argument, showed that  reliable communication in the sense of message recovery  is possible as long as 
  the information transmission rate  is below the channel capacity.
A fundamental  assumption in Shannon's model  is that the channel is known to the communicants.
For  Discrete Memoryless Channels (DMC) the probability law 
for $n$ times applications of the channel is given by,
$\mathsf{W}^n(\mathbf{y}|\mathbf{x}) =\Pi_{t=1}^n W(y_t|x_t) $ where  $W(y_t|x_t)$  is specified  by a probability stochastic matrix 
$W$ labeled by the elements of the input alphabet 
${\cal X}$ (rows)  and the output alphabet  ${\cal Y}$ (columns) , and 
$\mathbf{x}=(x_1\cdots x_n)  \in {\cal X}^n$ and $\mathbf{y}=(y_1\cdots y_n)\in {\cal Y}^n$.
\remove{

 the case of 
a Discrete Memoryless Channel (DMC), the channel is represented by a  transition probability matrix with rows and columns labelled by the input and output alphabets,
respectively,  and the $(i,j)^{th}$ entry is $\Pr(j|i)$, denoting the probability of  receiving $j$ when $i$ is sent.
}
Blackwell, Brieman and Thomasian~\cite{AVC}  initiated the study of  {\em Arbitrarily Varying Channels (AVC's)}: communication  channels with unknown parameters that can  vary  with time and in an unknown way. 
 A  {\em discrete memoryless AVC} (referred to simply as AVC in the sequel),  $\mathsf{W}:{\cal X}\times{\cal S} \rightarrow{\cal Y}$, is specified by
 a set of  
 stochastic matrices $\{\mathsf{W}_s\colon{\cal X}\rightarrow{\cal Y}|s\in {\cal S} \}$,  
 and
 ${\cal S}$ is called the set of {\em states} of the channel. 
 %
 For an input  sequence $\mathbf{x}=(x_1,\ldots,x_n)\in{\cal X}^n$ to the AVC,  the output distribution  is given by  $\mathsf{W}^n(\mathbf{y}|\mathbf{x}, \mathbf{s}) =\Pi_{t=1}^n W(y_t|x_t, s_t)$, where $s_t\in {\cal S}, t=1\cdots n$.
  AVCs  capture adversarially controlled channels, in particular, the set of channel states and their choice for each  symbol captures possible adversary's influence on transmitted symbols.
%
The traditional goal of communication over  AVC's is 
message recovery by the receiver. 
%
For powerful adversaries   however 
  one cannot expect  the message to be recoverable, and so 
   less demanding goals  such as detection of tempering, have been considered.
  Recently, Kosut and Kliewer \cite{KK18} studied     message authentication in a setting where 
the set of states includes a ``no adversary" state in which transmission will not be influenced by the adversary.
 This
 captures a  real life situation where 
  the adversary may or may not be present.
They considered 
two communication  goals: when the adversary is absent, 
 the error due to the stochastic matrix of the channel 
 that corresponds to the no adversary state must  
 be {\em corrected}; otherwise,  
  the presence of the adversary must be {\em detected}.  
  %
Codes for  detection of 
 tampering in  {\em  shared secret key setting} 
  was first considered by Gilbert, MacWilliams and Sloane \cite{GMS74}. They  introduced  Message Authentication Codes (MAC) that 
  can detect arbitrary tampering in the form of message substitution or injection of fruadulant messages.
  MAC is one of the most widely used cryptographic primitives for protection of communication.
\remove{
Recently, Kosut and Kliewer \cite{KK18} initiated the study of   message authentication in a special setting of AVC and without the need for a shared key.
They considered an AVC model with a {\em no adversary} state, which captures a real life situation that the adversary may or may not be present: when the adversary is present, the state can vary arbitrarily among the given set of states; when the adversary is not present, the state is fixed to the no adversary state.
They considered 
two communication  goals: when the adversary is absent, 
 the noise from the DMC corresponding to the no adversary state must  
 be {\em corrected}; otherwise,  
  the presence of the adversary must be {\em detected}.  
}
Kosut et al.'s model however does not assume shared key.

In this paper we consider a yet weaker goal for communication  for AVCs  
called {\em non-malleability}, 
 that prevents  the adversary from  tampering with the communication such that the decoded message be ``related" to the sent one.
{\em Non-malleability}  has been a widely studied goal in cryptography \cite{DDN00}. 
More recently Dziembowski, Pietrzak and Wichs \cite{DzPiWi} introduced {\em non-malleable codes (NM codes)}, 
where the adversary is defined by a family of tampering functions, $\cal F$, and non-malleable is defined as a relaxation of error correction and error detection.
 %
NM codes  
 are binary stochastic codes with randomized encoders and deterministic decoders. 
 Non-malleability is defined by the following experiment: 
a message $\mathbf{m}$ is encoded to a (randomized)  codeword; the adversary chooses 
a function $f \in \cal F$ and  tampers with the codeword;  and finally 
the (deterministic) decoder is used to decode the tampered codeword. 
Non-malleability requires that 
the decoded message is either the same as the 
encoded one, or results in a  
random message which is distributed according to a distribution that is  
determined by the tampering function $f$ only (and is independent of the encoded message). 

The following example elucidates the role of the function family in achieving non-malleability.
It is easy to see that 
 {\em linear} stochastic codes cannot provide  non-malleability 
 with respect to the family 
 ${\cal F} =\{f(\mathbf{x})=\mathbf{x}+\Delta|\Delta\in\{0,1\}^n\}$. 
 This is because  for an encoding $\mathbf{c}$ of a message $\mathbf{m}$,
 one 
 can choose  $\Delta$ to be 
 the codeword corresponding to the message $\mathbf{1}$, consisting of all 1's. 
  By applying the tampering function 
   $f(\mathbf{x})=\mathbf{x}+\Delta$ to  $\mathbf{c}$, we have $\mathbf{c}+\Delta$ which because of the linearity of the code,  is the encoding of  the 
   message $\mathbf{m} +\mathbf{1}$, which is equal to $\mathbf{m}$ with every  bit flipped,
   contradicting the definition of non-malleability. 

It has been shown that non-malleable codes exist  if 
 $\log\log|{\cal F}|\leq \alpha n$, for any $\alpha<1$ \cite{DzPiWi}. 
 Using a random coding argument, Cheraghchi and Guruswami \cite{ChGu0} derived the   lower bound  $1-\log\log|{\cal F}|/n$, on the achievable rate of NM codes for a  function family  ${\cal F}$. 
  This bound depends only on the size of the function family.
 %
An important   
 family of tampering functions is the Bit-wise Independent Tampering (BIT) family   in which 
 a function  $f\colon\{0,1\}^n\rightarrow\{0,1\}^n$ can be written as $f=(f_1,\ldots,f_n)$, where 
 $f_i\colon\{0,1\}\rightarrow\{0,1\}$ for $i=1,\ldots,n$.
  Explicit constructions of non-malleable codes for the BIT family that achieves information rate $1$, were given 
   in \cite{ChGu1,Maji}. 
   We will also use  non-malleable codes for 
   the family  of affine tampering functions  $f\colon\{0,1\}^n\rightarrow\{0,1\}^n$, where each output bit of a tampering function  
   can be  written as an affine function of the $n$ input bits. 
   Non-malleable codes for this family 
   were explicitly constructed in \cite{XinLi17}.

\smallskip
\noindent
\textbf {Our work.}
In this paper, we introduce non-malleability as the communication goal of 
(discrete memoryless)  AVCs. That is, 
 we require that  
the adversary cannot 
 modify the communication  in a way that the decoded message by the receiver be {\em related} to the sent one: 
  any tampering through the AVC will  
  result in either the   sent message to be correctly received, or completely lose its informational value.
%
We model the adversarial channel 
by 
an AVC $\{\mathsf{W}_s\colon{\cal X}\rightarrow{\cal Y}|s\in {\cal S} \}$, and require that for any state sequence $\mathbf{s}\in\mathcal{S}^n$,
the non-malleability guarantee should hold. 
 We focus on AVCs where the input and output alphabets  are  binary. 
 %
 
  We show that for such AVCs  
 non-malleable communication is always possible at rate  $1$. 
We show this  by proving that non-malleable codes for the family bitwise independent tampering can be used 
for   {\em non-malleable  coding} over these channels, 
achieving rate 
 $1$. 

We also consider a setting similar to Kosut and Kliewer \cite{KK18} where 
  there is a special state $s^{*}$, specified by the binary channel ${\mathsf{W}}_{s^{*}}$, for which we require the property that if used on all bits
   of the  sent codeword, allow the original message to  be recovered. The special state that we consider is defined by an erasure channel where each binary input symbol is erased with a constant probability $p^*$.   We thus expand the set of possible states of the AVC by the set of erasure channels.
\remove{
In other words if ${\mathsf{W}}_{s^{*}}$ is applied to all codeword bits,  
  the receiver is 
  can 
   recover the sent message.
   }
With this AVC with   the  new set of states, 
   we require the  communication to be  non-malleable.
We outline a 
construction that uses  a non-malleable code with respect to the family affine tampering functions and a linear erasure  correcting code, and provide the combined requirements.

\smallskip
\noindent
\textbf {Discussion and future work.}
We consider  non-malleable  coding for AVCs with binary alphabets (input and output).
To our knowledge non-malleability is the  least demanding goal for  protecting  a communication channel against tampering. 
Binary discrete memoryless  AVCs that  are  considered in this paper are a natural starting point for the study of NM channel codes.
This is  inline with the study of bitwise tampering function family 
 in \cite{DzPiWi} that first introduced NM codes. 
%
%
The power of the adversary in (discrete memoryless)  AVCs  is modelled by the set of possible states.
%
%
 An interesting research  question is to capture realistic tampering adversaries for communication channels as 
 AVCs, and design  non-malleable channel codes for them.
 \remove{
 define function families 
 that capture realistic tampering with communication channels 
 }
In  \cite{ISIT}, non-malleable codes for a  family functions  that  is inspired by 
 the adversarial  tampering of a communication channel is considered:  the adversary chooses the  
 tampering functions after observing some of the codeword bits (adversary chooses which bit to observe). 
Considering a similar function family in AVC setting requires more general AVCs that allow the state of the channel for a particular symbol to depend on the adversary's observation of the channel. 
\remove{
A similar type of adversarial access  
can be considered for  AVC setting 
where the channel state sequence is chosen by the adversary after observing a fraction of codeword bits. 
}
Other function families that have been considered for non-malleable codes    could  also be considered in AVC setting.

\remove{
For example in constant split state model,   an  $n$-bit codeword is split into a constant  number, $C$,  of blocks and each block 
is  independently tampered with. This could be formulated as AVCs over $q$-ary alphabets, where $q=2^{n/C}$. 
Extending our work to such a setting is an interesting open question.
%
}

\section{Models and main results} \label{sec: model}
We use the following notations.  A sequence (vector) of $n$ elements $x_i\in {\cal X}$ is  denoted by a bold
symbol: for example  ${\bf x} =(x_1\cdots x_n)$. 
%
The \emph{statistical distance} (total variation distance) between two  random variables  (their corresponding
 distributions) $\mathsf{X}$ and $\mathsf{Y}$ that are defined over the set 
 $\Omega$,   is defined as follows. 
$$\mathsf{SD}(\mathsf{X};\mathsf{Y})= \dfrac{1}{2}\sum_{\mathbf{\omega} \in \Omega}|\mbox{Pr}[\mathsf{X}=\mathbf{\omega}]-\mbox{Pr}[\mathsf{Y}=\mathbf{\omega}]|.$$
We say $\mathsf{X}$ and $\mathsf{Y}$ are $\varepsilon$-close (denoted $\mathsf{X}\stackrel{\varepsilon}{\approx}\mathsf{Y}$) if $\mathsf{SD}(\mathsf{X};\mathsf{Y})\leq \varepsilon$.

\subsection{Non-malleable coding for AVC}
We consider a discrete memoryless AVC, $\mathsf{W}:{\cal X}\times{\cal S} \rightarrow{\cal Y}$, with  
(discrete) input and  output alphabet sets $\X$ and $\Y$, respectively, and
$\mathcal{S}$ denoting the set of possible states. 
\remove{
Let $\mathsf{W}(y|x,s)$ be the probability that the input $x\in\mathcal{X}$ is mapped to $y\in\mathcal{Y}$ when the state is $s\in\mathcal{S}$.
 For a  sequence $\mathbf{s}\in \mathcal{S}^n$, an input sequence $\mathbf{x}\in\mathcal{X}^n$ and an 
output 
sequence $\mathsf{y}\in\mathcal{Y}^n$,  
the channel  transition probability is given by
$$
\mathsf{W}(\mathbf{y}|\mathbf{x},\mathbf{s})\colon=\prod_{i=1}^{n}\mathsf{W}(y_i|x_i,s_i).
$$
}
The set of states is 
corresponding to the set of stochastic matrices
$\{\mathsf{W}_s\colon{\cal X}\rightarrow{\cal Y}|
s\in {\cal S} \}$.
\remove{
 The probability that the input $x\in\mathcal{X}$ is mapped to the output $y\in\mathcal{Y}$ by $\mathsf{W}_s$ is given by $\mathsf{W}_s(y|x)=\mathsf{W}(y|x,s)$.
With the state  sequence fixed to $\mathbf{s}=(s_1,\ldots,s_n)\in \mathcal{S}^n$, 
}
The application of $\mathsf{W}$ to an input sequence $\mathbf{x}=(x_1,\ldots,x_n)\in\mathcal{X}^n$ for a state sequence   is given by,
$$
\mathsf{W}_{\mathbf{s}}(\mathbf{x})=(\mathsf{W}_{s_1}(x_1),\ldots,\mathsf{W}_{s_n}(x_n)).
$$
In this paper we consider 
AVC with binary input and output alphabet 
 $\X=\Y=\{0,1\}$. 

%


Let  $\bot$  be a special symbol that denotes 
 detection of tampering.  

\begin{definition}[\cite{DzPiWi}] A {\em $(k,n)$-coding scheme} consists of 
a randomized 
encoding function 
$\mathsf{Enc}:\{0,1\}^k\rightarrow\{0,1\}^n$ 
(randomness is implicit),  and a deterministic decoding function $\mathsf{Dec}:\{0,1\}^n\rightarrow\{0,1\}^k\cup\{\perp\}$ such that, for each $\mathbf{m}\in\{0,1\}^k$, $\mathsf{Pr}[\mathsf{Dec}(\mathsf{Enc}(\mathbf{m}))=\mathbf{m}]=1$, and the probability is  over the randomness of  encoding. 
\end{definition}

\remove{
 Let $s^0,s^1\in\mathcal{S}$ be two states defined by the 
  following channel transition matrix. 
$$
\begin{array}{l}
\mathsf{W}(0|0,s^0)=\mathsf{W}(0|1,s^0)=1; \mathsf{W}(1|0,s^0)=\mathsf{W}(1|1,s^0)=0.\\
\mathsf{W}(0|0,s^1)=\mathsf{W}(0|1,s^1)=0; \mathsf{W}(1|0,s^1)=\mathsf{W}(1|1,s^1)=1.
\end{array}
$$
We refer to these two channels as {\em 0-overwrite}  and  {\em 1-overwrite} channels, or constant channels, as they simply output the same value irrespective of the input.
Note that if  the set of states in an AVC includes these channels, and the coding scheme is public, the adversary can always overwrite the sent codeword with a codeword of its choice, making it impossible to have any detection  about the  decoded message.
%
%
That is the adversary overwrites the input codeword 
with a valid codeword $\tilde{\mathbf{c}}$ of a message $\tilde{\mathbf{m}}$. According to the definition of a coding scheme, the decoder will always output the message $\tilde{\mathbf{m}}$.  
}

A {\em tampering function} for a $(k,n)$-coding scheme is a function $f:\{0,1\}^n\rightarrow\{0,1\}^n$ that is applied to a codeword, and results in a binary $n$-vector.
Consider an experiment where a  message $\mathbf{m}$  is encoded by the encoder,    the codeword  is tampered by a function $f$, 
the decoder is   applied on the  tampered codeword  
outputting a message $\tilde{\mathbf{m}}$.
 Let $\mathsf{same}^{*}$ be a special symbol which means that $\tilde{\mathbf{m}}$ 
 is the same as $\mathbf{m}$. 
\begin{definition}[\cite{DzPiWi}] \label{def: non-malleability}Let $\mathcal{F}$ be a family of tampering functions.
For each $f\in\mathcal{F}$ and $\mathbf{m}\in\{0,1\}^k$, define the tampering-experiment
$$
\mathrm{Tamper}_\mathbf{m}^f=\left\{
\begin{array}{c}
\mathbf{x}\leftarrow\mathsf{Enc}(\mathbf{m}),\tilde{\mathbf{x}}= f(\mathbf{x}),\tilde{\mathbf{m}}=\mathsf{Dec}(\tilde{\mathbf{x}})\\
\mathrm{Output}\ \tilde{\mathbf{m}}\\
\end{array}
\right\},
$$
which is a random variable over the set $\{0,1\}^k\cup\{\bot\}$, and the randomness is due to 
the randomized encoding. 
A coding scheme $(\mathsf{Enc},\mathsf{Dec})$ is  {\em non-malleable with respect to $\mathcal{F}$}  if for any  $f\in\mathcal{F}$, there exists a distribution
 $\mathcal{D}_f$ over the set $\{0,1\}^k\bigcup\{\perp, \mathsf{same}^*\}$ such that, for all $\mathbf{m}\in\{0,1\}^k$, we have:
$$
\mathrm{Tamper}_\mathbf{m}^f\stackrel{\varepsilon}{\approx}\left\{
\begin{array}{c}
\tilde{\mathbf{m}}\leftarrow\mathcal{D}_f\\
\mathrm{Output}\ \mathbf{m}\ \mathrm{ if }\ \tilde{\mathbf{m}}=\mathsf{same}^*;\ \tilde{\mathbf{m}}\ \mathrm{ otherwise.}
\end{array}
\right\}, 
$$
and $\mathcal{D}_f$ is  efficiently samplable. 
\end{definition}

That is, $\mathrm{Tamper}_\mathbf{m}^f$ is a random variable that is  $\epsilon$-close to a distribution that is defined by the 
right hand side above, which  
is commonly  denoted  by $\mbox{Copy}(\mathcal{D}_f,\mathbf{m})$. 
Non-malleability is then rewritten as follows.
\begin{equation} \label{nmdef}
\mathrm{Tamper}_\mathbf{m}^f\stackrel{\varepsilon}{\approx}\mbox{Copy}(\mathcal{D}_f,\mathbf{m}). 
\end{equation}

Our definition of  {\em non-malleable codes for AVCs } can be seen as a natural extension of non-malleable codes to include probabilistic tampering.

\begin{definition}\label{def: AVC} Let $\mathsf{W}:{\cal X}\times{\cal S} \rightarrow{\cal Y}$ be a binary AVC. For a state sequence $\mathbf{s}\in\mathcal{S}^n$ and message $\mathbf{m}\in\{0,1\}^k$, consider the random variable $\mathrm{Tamper}_\mathbf{m}^{\mathsf{W}_\mathbf{s}}$ that is defined by the following tampering-experiment:
$$
\mathrm{Tamper}_\mathbf{m}^{\mathsf{W}_\mathbf{s}}=\left\{
\begin{array}{c}
\mathbf{x}\leftarrow\mathsf{Enc}(\mathbf{m}),\mathbf{y}\leftarrow \mathsf{W}_\mathbf{s}(\mathbf{x}),\tilde{\mathbf{m}}=\mathsf{Dec}(\mathbf{y})\\
\mathrm{Output}\ \tilde{\mathbf{m}}.\\
\end{array}
\right\}.
$$
This  random variable is  over the set $\{0,1\}^k\cup\{\bot\}$, and 
the randomness is from the encoder and also the application of channel 
$\mathsf{W}_\mathbf{s}(\cdot)$.
The coding scheme is $\varepsilon$-non-malleable for the AVC,  if for any message $\mathbf{m}\in\{0,1\}^k$ and any state sequence $\mathbf{s}\in\mathcal{S}^n$, there exists a distribution $\mathcal{D}_\mathbf{s}$ over the set $\{0,1\}^k\cup\{\bot,\mathsf{same}^{*}\}$ satisfying
$$
\mathsf{SD}(\mathrm{Tamper}_\mathbf{m}^{\mathsf{W}_\mathbf{s}};\mathsf{Copy}(\mathcal{D}_\mathbf{s},\mathbf{m}))\leq\varepsilon,
$$
where the distribution $\mathcal{D}_\mathbf{s}$ is independent of the message $\mathbf{m}$.
\end{definition}





Our main result is to show that   non-malleability for  binary AVCs  is  always  achievable.
We show this by 
proving that a  non-malleable code with respect to the BIT function family,  
 provides non-malleability for  transmission over 
  binary AVCs. 
%


\begin{theorem}\label{th: BIT} 
A $\varepsilon$-non-malleable coding scheme for 
BIT function family is a $\varepsilon$-non-malleable coding scheme for a binary AVC $\mathsf{W}:\{0,1\}\times \mathcal{S}\rightarrow\{0,1\}$. 
\end{theorem}

\subsection{Application to AVC with a special state}
We next consider a setting where one of the states $s^{*}\in \cal S$ is a special state,   in the sense that if that state is chosen for all $ 1\leq i\leq n$,  the receiver will  be able to recover the message.  
That is in addition to non-malleability that is guaranteed for any state sequence, we require the additional guarantee of message recovery for a special  state sequence $(s^{*})^n$. This special sequence captures a known state of the channel by the sender and the receiver and their goal of providing reliable communication for that. 

\begin{definition}\label{def: AVC'} Let $\mathsf{W}:{\cal X}\times{\cal S} \rightarrow{\cal Y}$ be a binary AVC with a special state $s^{*}\in\mathcal{S}$.  A $(k,n)$-coding scheme is $(\delta,\varepsilon)$-non-malleable with respect to $\mathsf{W}$, if the following properties hold.
\begin{enumerate}
\item For any message $\mathbf{m}\in\{0,1\}^k$, when the state sequence $\mathbf{s}=(s^{*})^n$, 
$$
\mathsf{Pr}[\mathsf{Dec}(\mathsf{W}_\mathbf{s}(\mathsf{Enc}(\mathbf{m})))=\mathbf{m}]\geq 1-\delta.
$$
\item For any message $\mathbf{m}\in\{0,1\}^k$ and any state sequence $\mathbf{s}\in (\mathcal{S}^n\setminus \{(s^{*})^n\})$, there exists a distribution $\mathcal{D}_\mathbf{s}$ over the set $\{0,1\}^k\cup\{\bot,\mathsf{same}^{*}\}$ satisfying
$$
\mathsf{SD}(\mathrm{Tamper}_\mathbf{m}^{\mathsf{W}_\mathbf{s}};\mathsf{Copy}(\mathcal{D}_\mathbf{s},\mathbf{m}))\leq\varepsilon,
$$
where the distribution $\mathcal{D}_\mathbf{s}$ is independent of $\mathbf{m}$. 
\end{enumerate}
\end{definition}




The special state  that we consider corresponds to a Binary Erasure Channel (BEC)
. We thus extend the set of states of the AVC $\mathsf{W}:\{0,1\}\times{\cal S} \rightarrow\{0,1\}\cup\{\bot\}$ to include 
BEC's, and use a specific BEC $\mathsf{W}_{s^{*}}$ for the special state $s^{*}\in\mathcal{S}$.

We sketch a  
generic
 construction that uses  
  a non-malleable code for the family of {\em affine} tampering functions and a linear erasure 
correcting code. 
The construction first 
encodes a $k$ bit message using  
a $(k,m)$ non-malleable coding scheme for the family of affine functions on $m$ bits, 
 and then encodes the resulting 
$m$-bit codeword using an erasure 
 correcting code into a $n$-bit final codeword for the BEC $\mathsf{W}_{s^{*}}$.  We use a decoder that correctly decodes the message if there are up to $p^*n$ erasures, where $p^*$ is erasure probability of  $\mathsf{W}_{s^{*}}$, and  declares failure (error detection) for more erasures. 
  When the state sequence  is $\mathbf{s}=(s^{*})^n$, the erasure 
   correcting code guarantees that the message is correctly recovered with probability at least $1-\delta$, where $\delta$ can be calculated for the code. 
For an arbitrary state sequence for the AVC that includes the erasure channels and  
 $\mathbf{s}\neq(s^{*})^n$, the proof  intuitively works as follows. 
 If the  received word has too many erasures, the decoder will  flag  detection.  When the number of erasures is less than 
 $p^*n$, the decoder will recover an $n$ bit string.  As will be shown in Section IV, the effect of this decoding  is that tampering on the $m$-bit NM codeword will be an affine function, 
 that can be protected against because of the property of the NM code with protection against affine tampering.



\section{Proofs of theorems} \label{sec: bounds} 

Let  $\bt$ denote the family of  Bitwise Independent Tampering (BIT) functions. 
For a  binary string 
of  length $n$, a BIT function $f\in\bt$ is written as $f=(f_1,\ldots,f_n)\in\{ \mathsf{Keep}, \mathsf{Flip}, \mathsf{Set0}, \mathsf{Set1}\}^n$, 
where $\mathsf{Set0}$ and $ \mathsf{Set1}$ overwrite the input with $0$ and $1$, respectively, and  $ \mathsf{Keep}$ and $ \mathsf{Flip} $ are the identity (no change) and the flip functions.

Non-malleable coding schemes for the function family  $\mathcal{F}_\mathsf{BIT}$ has been widely studied. In fact, the fist  construction of non-malleable codes \cite{DzPiWi} was given for this family.
The following theorem shows that an NM code that provides protection against the function family $\mathcal{F}_\mathsf{BIT}$ will provide protection against any binary AVC. 

\smallskip 
\noindent
{\em Proof of Theorem \ref{th: BIT}.}
Define the  set of {\em elementary binary channels} to be  
  $\{\mathsf{W}_{e_1}, \mathsf{W}_{e_2},\mathsf{W}_{e_3}, \mathsf{W}_{e_4}\}$. 
%
 corresponding to the bit functions $\{\mathsf{Keep}, \mathsf{Flip}, \mathsf{Set0}, \mathsf{Set1}\}$. 
The proof has the following steps. 
First, we show that a binary channel can be written as 
a convex combination of the  {\em  four elementary binary channels}.
Second,  using the above result 
we show that a sequence of   binary channels  of length $n$ (that is applied to an input sequence of length $n$) 
can be written as 
a convex combination of the $4^n$   channel sequences of length $n$ over elementary channels.
Finally, we use the properties of non-malleable codes 
for $\mathcal{F}_\mathsf{BIT}$ to show 
non-malleabilityis achievable  for  any binary AVC.
More details are given below.

\noindent
{\bf Claim 1:} {\em A  binary channel 
 can be written as a convex combination of four elementary channels. }\\
We use $\mathsf{W}_s$ to denote a binary channel and its two-by-two channel transition matrix. 
 First, consider a  binary DMC,  with channel transition matrix
$$
\mathsf{W}_s=\left [
\begin{array}{cc}
w_{11}&w_{12}\\
w_{21}&w_{22}\\
\end{array}
\right ],
$$
where rows and columns are  labeled by possible inputs and outputs, respectively.
The matrix entries  $w_{ij}$'s  are $ \mathsf{W}_s(j-1|i-1), i,j\in \{1,2\}$  and
satisfy the following:
\begin{equation} \label{eq: channel}
 0\leq w_{ij}\leq 1\;\; \; \mbox{ and,}\;\; \sum_j w_{ij}=1, \;\; i=1,2.
 \end{equation}
\remove{Last minute
 Similarly we have 
 $$
 \mathsf{W}_{e_1}=\left [
\begin{array}{cc}
1&0\\
0&1\\
\end{array}
\right ],\ 
\mathsf{W}_{e_2}=\left [
\begin{array}{cc}
0&1\\
1&0\\
\end{array}
\right ],
$$

$$
\mathsf{W}_{e_3}=\left [
\begin{array}{cc}
1&0\\
1&0\\
\end{array}
\right ],\ 
\mathsf{W}_{e_4}=\left [
\begin{array}{cc}
0&1\\
0&1\\
\end{array}
\right ].
 $$
 }
We next show the matrix $\mathsf{W}_s$ can be written as a sum of the transition matrices of the four elementary channels: 
\begin{equation} \label{eq: convex}
\mathsf{W}_s= \alpha_1\mathsf{W}_{e_1}+\alpha_2 \mathsf{W}_{e_2}+\alpha_3 \mathsf{W}_{e_3}+\alpha_4 \mathsf{W}_{e_4}
\end{equation}
where the 
coefficients $\alpha_1, \alpha_2, \alpha_3, \alpha_4$  are non-negative real numbers that satisfy 
$\alpha_1+\alpha_2+\alpha_3+\alpha_4=1$.  
\remove{Last minute
To find  these coefficients  using 
(\ref{eq: channel}) and (\ref{eq: convex}), we  need to solve 
the following linear system
\begin{eqnarray*}
\alpha_1+\alpha_3=w_{11}, \;\;
\alpha_2+\alpha_4=1-w_{11}\\
\alpha_2+\alpha_3=1-w_{22},\;\;
\alpha_1+\alpha_4=w_{22}\\
\end{eqnarray*}
}
By solving linear equations, we have the following relations, 
\begin{eqnarray*}
&&\alpha_1=w_{11}-\alpha_3, \;\;
\alpha_2=1-w_{22}-\alpha_3, \\
&&
\alpha_4=\alpha_3-(w_{11}-w_{22})
\end{eqnarray*}
Note that we do have $\alpha_1+\alpha_3+\alpha_2+\alpha_4=1$.
What is left to show 
 is that these coefficients can be chosen to be all non-negative. This can be satisfied if 
 the following holds,
\[1-w_{22}\geq\alpha_3 \geq 0,  \mbox{ and }\;\; w_{11}\geq\alpha_3\geq w_{11}-w_{22}.\]
When $w_{11}\geq w_{22}$, we can let $\min\{w_{11},1-w_{22}\}\geq\alpha_3\geq w_{11}-w_{22}$. When $w_{11}<w_{22}$, we let $\min\{w_{11},1-w_{22}\}\geq\alpha_3\geq0$.  Thus 
 such coefficients always exist, and so 
 choosing the value of $\alpha_3$  allows computation of $\alpha_1, \alpha_2$ and $ \alpha_4$.



\noindent
{\bf Claim 2: } 
{\em
 For 
a state sequence of length $n$, the corresponding 
application  of the 
The output of a binary AVC  of length $n$ with a state  sequence  $\mathbf{s}$ on an input sequence $\mathbf{x}$ of length $n$,  
can be written as a convex combination of the outputs of 
 $4^n$  AVCs,  each applied on the same input  $ \mathbf{x}$. The AVCs  correspond  to  the $4^n$ possible state sequences of length $n$ over elementary   channels.  

\remove{
 For 
a state sequence of length $n$, the corresponding 
application  of the 
binary AVC on an input sequence $\mathbf{x}$ of length $n$  
can be written as a convex combination of the output of 
the $4^n$  elementary binary channel sequences 
of length $n$ on   $ \mathbf{x}$.  
}
}

Using the notations 
in Section \ref{sec: model}-A, for a state sequence $(s_1,\ldots,s_n)\in\mathcal{S}^n$, we have 
$$
\mathsf{W}_{\mathbf{s}}(\mathbf{x})=(\mathsf{W}_{s_1}(x_1),\ldots,\mathsf{W}_{s_n}(x_n)).
$$
Using the results of Claim 1, we have the following decomposition for $i=1,\ldots,n$, 
\begin{equation}\label{eq: composition}
\mathsf{W}_{s_i}= \alpha_{i,1}\mathsf{W}_{e_1}+\alpha_{i,2} \mathsf{W}_{e_2}+\alpha_{i,3} \mathsf{W}_{e_3}+\alpha_{i,4} \mathsf{W}_{e_4},
\end{equation}
where  $ \alpha_{i,j}$ are non-negative real numbers and 
$$\sum_{j=1}^4  \alpha_{i,j } =1.$$
This means that applying  channel $\mathsf{W}_{s_i}$ to $x_i$  results in the same output distribution as   applying the elementary channels $\mathsf{W}_{e_j}$ with associated  probabilities $\alpha_{i,j}$. 
For the channel sequence $(\mathsf{W}_{s_1},\ldots,\mathsf{W}_{s_n})$, the probability that $\mathsf{W}_{e_{j_i}}$ is applied to $x_i$ for all $i=1,\ldots,n$ is then $\prod_{i=1}^n\alpha_{i,j_i}$.
Using $\mathsf{W}_{\mathbf{s}}$ to denote 
the channel sequence,  and its corresponding sequence of channel transition 
matrices, 
we have: 
$$
\begin{array}{ll}
\mathsf{W}_{\mathbf{s}}&=(\mathsf{W}_{s_1},\ldots,\mathsf{W}_{s_n})\\
                                       &=\sum_{(j_1,\ldots,j_n)\in[4]^n}\left(\prod_{i=1}^n\alpha_{i,j_i}\right) (\mathsf{W}_{e_{j_1}},\ldots,\mathsf{W}_{e_{j_n}}).
\end{array}
$$
The coefficients in the above decomposition are non-negative and we have
$$
\sum_{(j_1,\ldots,j_n)\in[4]^n}\left(\prod_{i=1}^n\alpha_{i,j_i}\right)=\prod_{i=1}^n\left(\sum_{j=1}^4  \alpha_{i,j }\right)=1.
$$

\noindent
{\bf Claim 3: }  {\em  An $\epsilon$-non-malleable code for function family   $\bt$  provides $\epsilon$-non-malleable protection for 
an AVC $\{0,1\}\times \mathcal{S}\rightarrow\{0,1\}$.}\\ 
Given a $(k,n)$-coding scheme that is $\varepsilon$-non-malleable with respect to the function family $f\in\{\mathsf{Keep}, \mathsf{Flip},\mathsf{Set0}, \mathsf{Set1}\}^n$, we want to show that for any state sequence $\mathbf{s}=(s_1,\ldots,s_n)\in\mathcal{S}^n$ there is a distribution $\mathcal{D}_\mathbf{s}$ that satisfies
$$
\mathsf{SD}(\mathrm{Tamper}_\mathbf{m}^{\mathsf{W}_\mathbf{s}};\mathsf{Copy}(\mathcal{D}_\mathbf{s},\mathbf{m}))\leq\varepsilon.
$$
\remove{
Recall that 
???we use the two sets of notations $\{\mathsf{W}_{e_1}, \mathsf{W}_{e_2},\mathsf{W}_{e_3}, \mathsf{W}_{e_4}\}$ and $\{\mathsf{Keep}, \mathsf{Flip},\mathsf{Set0}, \mathsf{Set1}\}$ ??? interchangeably. ??? WHAT IS THE MEANING OF THIS? REMOVE
}
Define the following projection maps for binary channels $\mathsf{W}_{s_i}$.
$$
\begin{array}{ll}
\Phi_{\mathsf{Keep}}(\mathsf{W}_{s_i})=\alpha_{i,1}, 
&\Phi_{\mathsf{Flip}}(\mathsf{W}_{s_i})=\alpha_{i,2},\\
\Phi_{\mathsf{Set0}}(\mathsf{W}_{s_i})=\alpha_{i,3},
&\Phi_{\mathsf{Set1}}(\mathsf{W}_{s_i})=\alpha_{i,4},\\
\end{array}
$$
where $\alpha_{i,1},\alpha_{i,2},\alpha_{i,3},\alpha_{i,4}$ are given in (\ref{eq: composition}).
According to Claim 2, we have that 
$$
(\mathsf{W}_{s_1},\ldots,\mathsf{W}_{s_n})=\sum_{f=(f_1,\ldots,f_n)\in\mathcal{F}_\mathsf{BIT}
}\left(\prod_{i\in[n]}\Phi_{f_i}(\mathsf{W}_{s_i})\right) f.
$$
Recall that the 
coding scheme is  $\varepsilon$-non-malleable with respect to  functions $f\in\{\mathsf{Keep}, \mathsf{Flip},\mathsf{Set0}, \mathsf{Set1}\}^n$. 
That is for the  function $f$, there exists 
 a distribution $\mathcal{D}_f$ 
  that satisfies 
$$
\mathsf{SD}(\mathrm{Tamper}_\mathbf{m}^f;\mathsf{Copy}(\mathcal{D}_f,\mathbf{m}))\leq\varepsilon.
$$
We  define
$$
\mathcal{D}_\mathbf{s}=\sum_{f\in\mathcal{F}_\mathsf{BIT}}\left(\prod_{i\in[n]}\Phi_{f_i}(\mathsf{W}_{s_i})\right)\mathcal{D}_f.
$$
Then $\mathcal{D}_\mathbf{s}$ is the corresponding distribution for the $\mathsf{W}_\mathbf{s}$ that satisfies Definition \ref{def: AVC}.

\smallskip

There are explicit constructions of non-malleable codes with respect to bit-wise independent tampering that achieves information rate $1$ \cite{ChGu1,Maji}. According to Theorem \ref{th: BIT}, we then have rate $1$ non-malleable codes for any AVC.

\begin{corollary} There exist non-malleable coding schemes for any AVC $\mathsf{W}:\{0,1\}\times{\cal S} \rightarrow\{0,1\}$ achieving rate $1$.
\end{corollary}

\section{A coding scheme for AVC with special state}


We  
consider a linear erasure correcting code 
for the BEC $\mathsf{W}_{s^{*}}$, with  encoder $\mathsf{ECCenc}\colon\{0,1\}^m\rightarrow\{0,1\}^n$ and the following decoder $\mathsf{ECCdec}\colon\{0,1\}^n\rightarrow\{0,1\}^m\cup\{\bot\}$. 

$\mathsf{ECCdec}$: \textit{
Let $G_{m\times n}$ be the generator matrix of the code. 
Consider a received word $\mathbf{y}$ with  erased bits  given by $E\subset[n]$.
The decoder  $\mathsf{ECCdec}$ finds a subset  $R\subset [n]\backslash E$  with $|R|=m$, such that $G_{R}$ is an invertible 
a submatrix of  $G_{m\times n}$ with columns corresponding to $R$.
Such submatrix will exist with overwhelming probability because of erasure correction property of the code. The decoder's output is,
\begin{equation} \label{decode}
\tilde{\mathbf{m}}=\mathbf{y}_RG_R^{-1}.
\end{equation}
If the set $E$ of erased bits is too large such that no reconstruction set $R$ exists, the decoder simply outputs $\bot$. 
}

As described in Section II-B, the construction uses a two step coding. We first define the notion of 
{\em induced tampering}. 
\remove{
Consider a linear erasure correcting code with a deterministic encoder that  maps a message  to 
a codeword,  and a deterministic decoder that maps a received word to a message.
An erasure correcting code must  satisfy the correctness property $\mathsf{ECCdec}\circ \mathsf{ECCenc}=\mathsf{Id}$, where $\mathsf{Id}$ denotes the identity function.
}
\begin{definition}[\cite{inducedtampering}]\label{def: induced tampering} Let $\mathsf{ECC}$ be an erasure 
correcting code with an encoder $\mathsf{ECCenc}\colon\{0,1\}^{m}\rightarrow\{0,1\}^n$ and a decoder 
 $\mathsf{ECCdec}$. 
Let $\mathcal{F}$ be a family of tampering functions over $\{0,1\}^n$. 
The tampering family $\mathcal{G}$ over $\{0,1\}^m$ induced by 
$\mathcal{F}$ and the ECC 
is defined as follows.
\begin{equation}\label{eq: induced}
\mathcal{G}\colon=\{\mathsf{ECCdec}\circ f\circ \mathsf{ECCenc}|f\in\mathcal{F}\},
\end{equation}
where ``$\circ$'' denotes the composition of functions.
\end{definition}

\remove{
\smallskip
\noindent
\textcolor{blue}{
$\mathsf{ECCdec}$: \textit{
Let the generator matrix of the ECC be $G_{m\times n}$. For $R\subset[n]$, let $M_{R}$ denote the submatrix of $M_{a\times n}$ consist of the columns of $M_{a\times n}$ specified by $R$, where $a$ is an arbitrary integer. Given a received word $\mathbf{y}$ with the bits in $E\subset[n]$ erased. $\mathsf{ECCdec}$ finds an $R\subset [n]\backslash E$ with $|R|=m$ such that $G_{R}$ is invertible and compute 
$$\tilde{\mathbf{m}}=\mathbf{y}_RG_R^{-1}.$$
If the set $E$ of erased bits is too large such that no reconstruction set $R$ exists, the decoder simply outputs $\bot$. 
}}
}
Let $(\mathsf{NMCenc},\mathsf{NMCdec})$ be a coding scheme that is non-malleable with respect to the family of affine tampering functions over $\{0,1\}^m$. 
Consider the $(k,n)$-coding scheme 
\begin{equation}\label{eq: construction}
\left\{
\begin{array}{ll}
\mathsf{Enc}(\mathbf{m})&=\mathsf{ECCenc}(\mathsf{NMCenc}(\mathbf{m}))\\
\mathsf{Dec}(\mathbf{y})&=\mathsf{NMCdec}(\mathsf{ECCdec}(\mathbf{y})).
\end{array}
\right.
\end{equation}

We will use an erasure correcting code with detection (failure) error $\delta$  that can recover  up to $p^*n$ errors.
Consider the case where $\mathbf{s}\neq (s^{*})^n$. If  the number of erased symbols less than $p^*n$
 (i.e., the adversary has not erased too many symbols), 
 the decoder will have an output.
 \remove{
We now show the existence of a distribution $\mathcal{D}_\mathbf{s}$ that satisfies the Definition \ref{def: AVC} for 
all such state sequences. 
 Using the argument in 
 }
 From the proof of Theorem \ref{th: BIT}, we only need to consider the state sequences that consists of 
  bit-wise independent tampering functions.  
For any bit-wise independent tampering function $f$ over $\{0,1\}^n$, let $g_f=\mathsf{ECCdec}\circ f\circ\mathsf{ECCenc}$ be the tampering function over $\{0,1\}^m$ induced by $f$, through the encoder and decoder of the erasure code. 
 
We now argue that $g_f$ is an affine function.  This follows   by noting that, (i) from equation \ref{decode} 
the decoder output $\tilde{\mathbf{m}} =\mathbf{y}_RG_R^{-1}$, and (ii) a  bitwise independent function $f=(f_1,\ldots,f_n)$  can be expressed as an affine function $f(\mathbf{x})=\mathbf{x}M_f+\Delta_f.$
\remove{
The non-erased  components of the received word are tampered by  the bitwise independent functions $f=(f_1,\ldots,f_m)$ which can be written as
\begin{equation}\label{eq: BITaffine}
f(\mathbf{x})=\mathbf{x}M_f+\Delta_f,
\end{equation}
The matrix $M_f$ and the offset $\Delta_f$  are obtained as follows. If $f_i=\mathsf{Keep}$, we set the $i$th component of $\Delta_f$ to $0$ and the $i$th column of $M_f$ to $\mathcal{E}_i$, the standard basis vector with the only non-zero entry $1$ at the $i$th position. If $f_i=\mathsf{Flip}$, we set the $i$th component of $\Delta_f$ to $1$ and the $i$th column of $M_f$ to $\mathcal{E}_i$. If $f_i=\mathsf{Set0}$, we set the $i$th component of $\Delta_f$ to $0$ and the $i$th column of $M_f$ to the zero column. If $f_i=\mathsf{Set1}$, we set the $i$th component of $\Delta_f$ to $1$ and the $i$th column of $M_f$ to the zero column.
}
The combination of these  result in  the induced function on the recovered word to be  given by,

\remove{

Since $f\in\bt$ is in particular an affine function, we have 
\begin{equation}\label{eq: BITaffine}
f(\mathbf{x})=\mathbf{x}M_f+\Delta_f,
\end{equation}
where $M_f$ is a $n\times n$ constant matrix determined by $f$ and $\Delta_f$ is a 
constant row vector determined by $f$.
\textcolor{red}{Indeed, for an arbitrary $f=(f_1,\ldots,f_n)$, we specify the matrix $M_f$ and the offset $\Delta_f$ as follows. If $f_i=\mathsf{Keep}$, we set the $i$th component of $\Delta_f$ to $0$ and the $i$th column of $M_f$ to $\mathcal{E}_i$, the standard basis vector with the only non-zero entry $1$ at the $i$th position. If $f_i=\mathsf{Flip}$, we set the $i$th component of $\Delta_f$ to $1$ and the $i$th column of $M_f$ to $\mathcal{E}_i$. If $f_i=\mathsf{Set0}$, we set the $i$th component of $\Delta_f$ to $0$ and the $i$th column of $M_f$ to the zero column. If $f_i=\mathsf{Set1}$, we set the $i$th component of $\Delta_f$ to $1$ and the $i$th column of $M_f$ to the zero column.
}

\remove{
Now let the generator matrix of the ECC be $G_{m\times n}$. We then have
$$
f\circ\mathsf{ECCenc}(\mathbf{u})=\mathbf{u}GM_f+\Delta_f,
$$
where $\mathbf{u}\in\{0,1\}^m$. 
For $R\subset[n]$, let $M_{R}$ denote the submatrix of $M_{a\times n}$ consist of the columns of $M$ specified by $R$. Given a received word with the bits in $E\subset[n]$ erased. The decoder of ECC finds an $R\subset [n]\backslash E$ such that $G_{R}$ is invertible and compute 
}
The tampering induced by $f\in\bt$ and the ECC is then either $g_f\equiv \bot$, when there are too many erasures, or
}
$$
g_f(\mathbf{u})=\mathsf{ECCdec}_R\circ f\circ\mathsf{ECCenc}(\mathbf{u})=(\mathbf{u}GM_f+\Delta_f)_R G_R^{-1},
$$
where 
$g_f$ is an affine function given any choice of $R\subset[n]$ with $|R|=m$ such that $G_{R}$ is invertible.



\section*{Acknowledgment}
The work of Safavi-Naini is supported in part by Natural Sciences and Engineering Research Council of Canada, Discovery Grant Program.
The work of Lin, Ling, Wang is supported by Singapore Ministry of Education under Research Grant MOE2016-T2-2-014(S) and RG133/17 (S).  

The authors list is  alphabetical.


 

\bibliographystyle{plain}
\bibliography{1.bib}

\end{document}